# Climate Network Structure Follows North Atlantic Oscillation Phases

O. Guez[1], A. Gozolchiani[1], K. Yamasaki[2], Y. Berezin[1], S. Brenner[3] and S. Havlin[1].

[1]   We construct a network from climate records of different geographical sites in the North Atlantic. A link between two sites represents the cross-correlations between the records of each site. We find that within the different phases of the North Atlantic Oscillation (NAO) the correlation values of the links are significantly different. By setting an optimize threshold on the correlation values, we find that the number of strong links in the network is increased during times of positive NAO indices, and decreased during times of negative NAO indices. We find a pronounced sensitivity of the network structure to the oscillations which is significantly higher compared to the observed response of spatial average of the records. Our result suggests a new measure that tracks the NAO pattern.

## 1. Introduction

[2]   A network approach has recently been applied in order to follow climate dynamics [*Tsonis et al.*, 2006. *Yamasaki et al.*, 2008]. The nodes of the climate network are geographical sites. The dynamics recorded in each site is composed of its intrinsic dynamics and to the coupling with the dynamics of other sites. The cross-correlations due to the coupling between the dynamics in two different sites are represented in our network by a link between the sites (see [*Castrejo`n-Pita et al.*, 2010] for a lab experiment that demonstrates the relation between the coupling and the correlation). The maximum value of the correlation might appear with a time-delay between the two data records. This approach has recently led to the discovery of several novel insights related to El-Nino dynamics [*Tsonis et al.*, 2008. *Yamasaki et al.*, 2008. *Tsonis*, 2008. *Gozolchiani et al*., 2008. *Tsonis*, 2007].

[3]   The North Atlantic Oscillation (NAO) is the dominant mode of winter climate variability on a decadal time scale in the North Atlantic region ranging from Central North America to Europe and into Northern Asia. The NAO is a large scale seesaw in atmospheric mass between the subtropical high and the polar low [*Van Loon et al*., 1978]. The corresponding index is measured by the difference between the mean winter sea level pressure over Azores and over Iceland [*Walker et al*., 1932]. It varies from year to year, although it exhibits a tendency to remain in one phase for intervals lasting several years [URL 1].

[4]   In this paper we concentrate on the collective behavior of the nodes and links of the climate network during the different NAO phases. In the second section we describe the methods we used, including the data and the numerical procedure. In the Third section we present our result, including also a comparison with a method that does not involve the second moment (correlations) and implications. In the last section we summarize and discuss the possible implications of our findings.

[1]Minerva Center and Department of Physics, Bar-Ilan University, Ramat-Gan 52900, Israel.
[2]Tokyo University of Information Sciences - Chiba, Japan.
[3]Department of Geography, Bar-Ilan University, Ramat-Gan 52900, Israel.

## 2. Methods

### 2.1 Data

[5] We analyze the global National Center for Environmental Prediction/National Center for Atmospheric Research (NCEP/NCAR) reanalysis air temperature field and geopontential height field, both at 1000hPa and 500hPa [*Kalnay et al.*, 1996]. For each node of the network, daily values for the period 1948-2006 are used, from which we extract anomaly values (actual values minus the climatological averaged over the years for each day).

[6] The data is arranged on a grid latitude-longitude with a resolution of $2.5° \times 2.5°$. The location on the globe of the studied region is shown in figure 1 where the NAO is known to affect [*Feldstein*, 2003], and includes the Icelandic-Low and the Azores-High pressure centers, where the NAO Index is measure [*Jones et al.*, 1997]. In this zone there are 33 grid nodes in the east-west direction and 25 grid nodes in the north-south direction, amounting to total of 825 grid points.

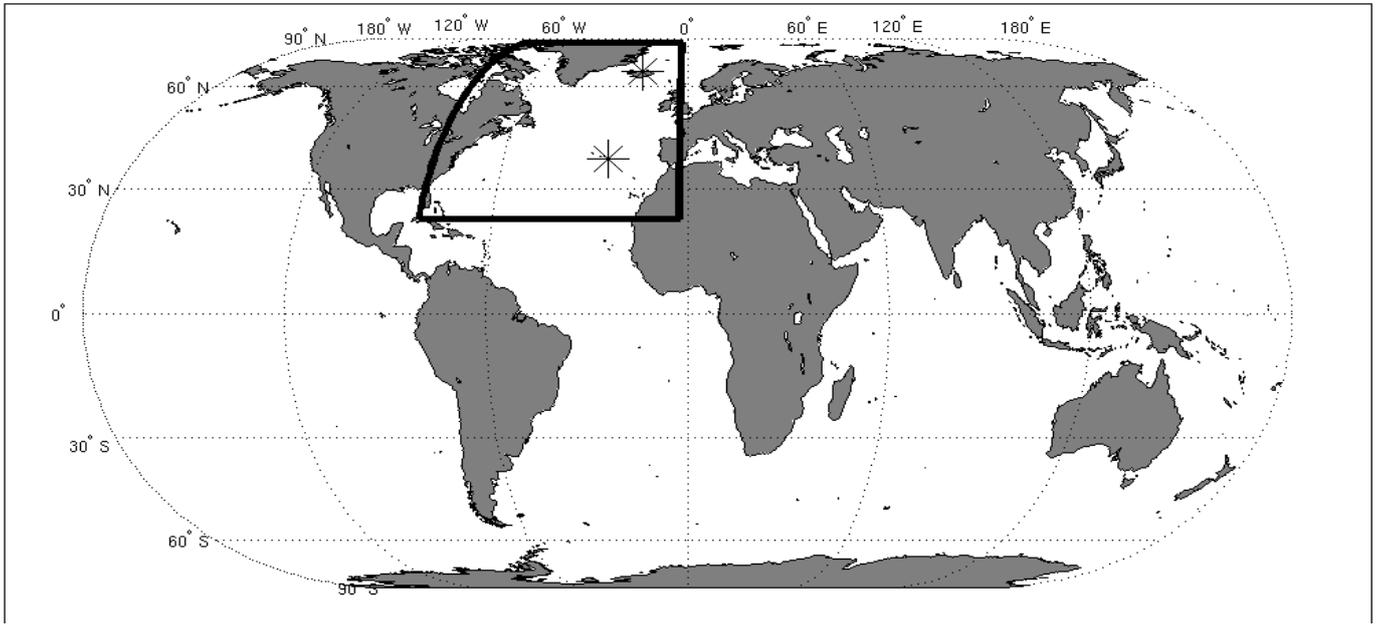

**Figure 1** The geographical region examine is $((22.5° - 82.5°)N, (2.5° - 82.5°)W)$. The Azores station is Ponta Delgada $(37.7°N, 25.7°W)$, shown as the upper star. The Iceland station is Akureyri $(65.7°N, 18.1°W)$, shown as the lower star [*Jones et al.*, 1997].

### 2.2 Numerical Procedure

[7] We first quantify the correlation value of the links, which characterizes the interdependence of the different pairs of sites. We compute for each pair of sites the Pearson correlation function (see e.g., [*Press et al.*, 1992]):

$$C_{M,N}^{Y}(\tau > 0) \equiv \frac{\left|\left\langle \left[D_{M}^{Y}(d) - \left\langle D_{M}^{Y}(d)\right\rangle\right] \cdot \left[D_{N}^{Y}(d+\tau) - \left\langle D_{N}^{Y}(d+\tau)\right\rangle\right]\right\rangle\right|}{\left\langle \left[D_{M}^{Y}(d) - \left\langle D_{M}^{Y}(d)\right\rangle\right]^{2}\right\rangle^{\frac{1}{2}} \cdot \left\langle \left[D_{N}^{Y}(d+\tau) - \left\langle D_{N}^{Y}(d+\tau)\right\rangle\right]^{2}\right\rangle^{\frac{1}{2}}}$$

(1)

$$C_{M,N}^{Y}(\tau<0) \equiv C_{N,M}^{Y}(\tau>0). \quad (2)$$

Here M and N are the indices of the two sites. D is the data record and d is the day index, ranging between 1 and 122 (1st December of the current year to 1st April of the following year). We chose to analyze only the winter season when the NAO effects are known to be stronger (see e.g., [URL1]). The parameter $\tau$ is the time lag, ranging between -72 and +72 days. Y is the year, ranging between 1948 and 2006.

[8] The Pearson correlations, $C_{M,N}^{Y}$, are computed from data that begins in the year represented by Y and ends in Y+2. This choice is made in order to have sufficient statistics as well as to capture the dynamical changes of these correlations. Upper limits of Y, Y+1, Y+3 and Y+4 yield similar results but less pronounced (not shown).

[9] To better quantify the interdependence of the different pairs of sites we define the strength of the link as the highest correlation value with respect to the background noise (see *Yamasaki et al.*, 2008). Thus, the strength of the link in year Y, connecting the sites M and N, $S_{M,N}^{Y}$, is define as follows:

$$S_{M,N}^{Y} = \frac{MAX(C_{M,N}^{Y}) - MEAN(C_{M,N}^{Y})}{STD(C_{M,N}^{Y})}. \quad (3)$$

The average (MEAN) and standard deviation (STD) are taken over all $\tau$ values ($-72 \leq \tau \leq +72$).

[10] To test whether and how the dynamics of these links is relate to NAO, we build a network composed of links stronger than a given threshold value, $H$. The existence (1) or nonexistence (0) of a strong link is represented in the time-dependent adjacency matrix of the network:

$$B_{M,N}^{Y}(H) = S_{M,N}^{Y} \cdot \Theta(S_{M,N}^{Y} - H) \quad (4)$$

Where $\Theta(x)$ is the unit step function (Heaviside).

[11] Summing $B_{M,N}^{Y}(H)$ over all pairs M,N for a given year Y yields the time-series of the number of the links in the network, for a given $H$:

$$\sigma_H(Y) = \sum_{M>N} B_{M,N}^{Y}(H). \quad (5)$$

## 3. Results

[12] In the following section we explore a relation between the structure of the climate network and the NAO events.

### 3.1 The Influence of the NAO on the Number of Links in the Network

[13] First we calculate the number of links, $\sigma_H(Y)$, that are stronger than threshold $H$. Then we compute the correlation $R(H)$ between $\sigma_H(Y)$ and the NAO index, $I(Y)$. To estimate the error in $R(H)$

we randomly divide the 339,000 pairs into 20 sub-networks. We perform the analysis described above for each sub-network. The standard-deviation of $R(H)$ of the twenty sub-networks is the estimator.

### 3.2 Optimization Method

[14] We systematically compute the correlation $R(H)$ for different $H$ values. The optimal threshold, $H^*$, was chosen as the one which yields the maximal correlation, $R(H^*)$. The same also preformed with a surrogate data (describe in [19]) and shown in figure 2 for each of the 4 data sets mention above. For data set (4) one can see that for $H < 5.2$` the correlations $R(H)$ of the real data are comparable to the correlations $R(H)$ of the surrogate data. However for higher values of $H$ there is a broad range of thresholds for which the correlation $R(H)$ of the real data are significantly higher than the correlations $R(H)$ of surrogate data. The value $H^* = 6.1$ is the optimal threshold, at which $R(H^*) = 0.50 \pm 0.01$ is the maximal correlation. Similar behavior is seen for the other data sets.

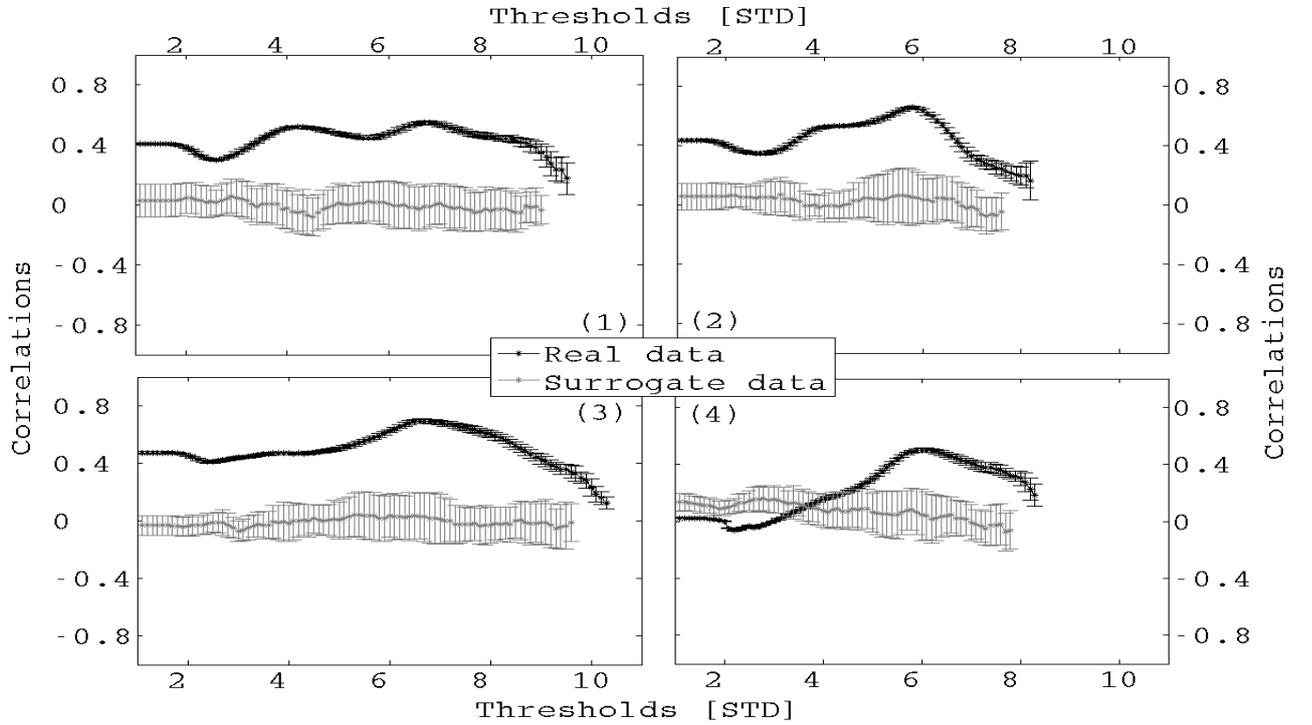

**Figure 2** The correlation, $R(H)$ (described in [14]), between the number of the strong links, $\sigma_H(Y)$ (described in [11]), and the NAO-index, $I(Y)$ (mentioned in [6]), as a function of the threshold, $H$ (described in [10]), for the real data (blue) and the surrogate data (green). The four panels describe the behavior of the network of four data sets: (1) Air temperature field at 500hPa. (2) Geopontential height field at 500hPa. (3) Air temperature field at 1000hPa. (4) Geopontential height field at 1000hPa.

### 3.3 Average Field Test

[15] We also calculate the correlation $\tilde{R}$ between the NAO-index, $I(Y)$, and the average of all the records:

$$\Delta(Y) \equiv \left\langle \left\langle D_M^Y(d) \right\rangle_d \right\rangle_M .$$
(6)

Here the sub index *d* represents averaging over the winter days and the outer sub-index M represents averaging over all sites. The error of the values of the correlations is estimated using the standard deviation of sub-networks (described in [13]).

[16] To compare the correlations $R(H^*)$ and $\tilde{R}$ we depict them in . One can see that for all data sets the absolute value of $\tilde{R}$ is significantly smaller than the value of $R(H^*)$. This indicates that the sensitivity of the number of strong links in the network to the NAO is higher compared with the sensitivity of the average of all the records.

| Data Set | $R(H^*)$ | $\tilde{R}$ |
|---|---|---|
| (1) | +0.55 ± 0.01 | -0.16 ± 0.09 |
| (2) | +0.66 ± 0.01 | -0.39 ± 0.09 |
| (3) | +0.70 ± 0.01 | -0.28 ± 0.12 |
| (4) | +0.50 ± 0.01 | -0.26 ± 0.01 |

**Table 1** The values of the correlations $R(H^*)$ (described in [14]) between the number of the strong links, $\sigma_H(Y)$ (described in [11]), and the NAO-index, $I(Y)$ (mentioned in [6]), at the optimal threshold, $H^*$ (mentioned in [14]). The values of the correlations $\tilde{R}$ (described in [15]) between the average of all the records, $\Delta(Y)$ (described in [15]), and the NAO-Index, $I(Y)$. The 4 lines show the correlations for 4 types of data sets: (1) Air temperature field at 500hPa. (2) Geopontential height field at 500hPa. (3) Air temperature field at 1000hPa. (4) Geopontential height field at 1000hPa.

### 3.4 Implications

[17] We complete this part with a summary of the implication of our calculations. We show in figure 3 the behavior of the number of the strong links, $\sigma_{H^*}(Y)$ (described in [11]), and average of all the records, $\Delta(Y)$, in comparison with the NAO-Index, $I(Y)$. One can see that $\sigma_{H^*}(Y)$ obtains high values when $I(Y)$ is positive and vice versa. This behavior suggests a positive correlation between the number of links in the network and the NAO-index (see also Table 1).

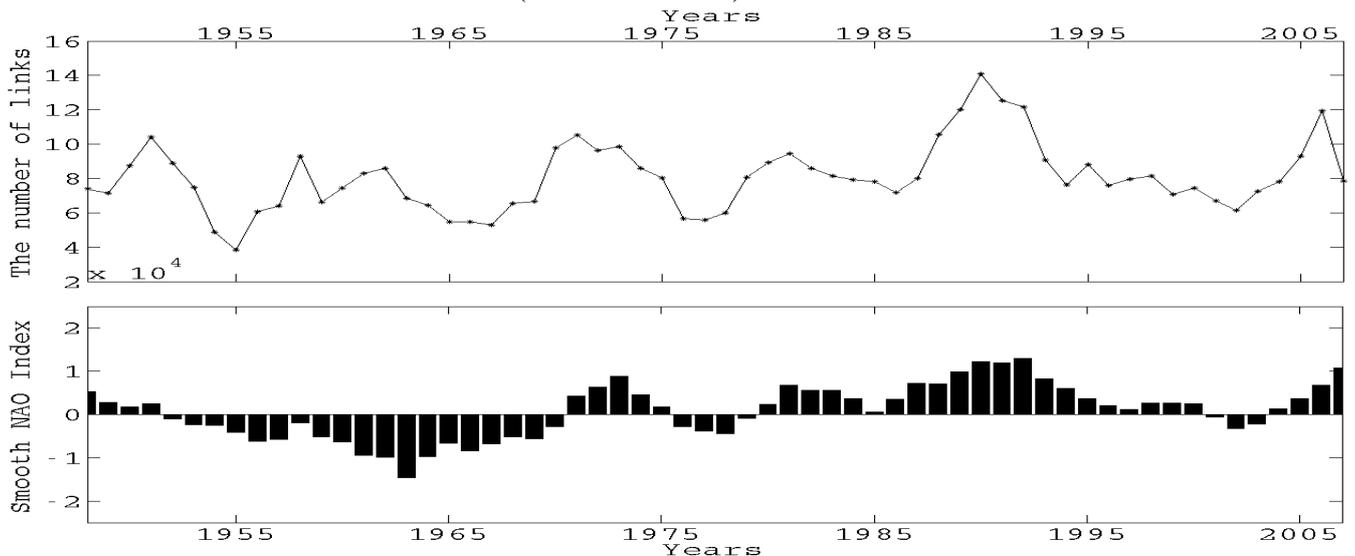

**Figure 3** The smooth NAO-Index, $I(Y)$ (lower plot, mentioned in [6]). The number of the strong links, $\sigma_{H^*}(Y)$ (upper plot, described in [11]), for the optimal threshold, $H^* = 6.8$ (mentioned in [14]), for data set (3) Air temperature field at 1000hPa.

## 4. Validation Tests

[18] Here we present two validation tests for the significance of our result for the correlations of the real data, $R(H)$.

[19] The first validation test is performed by using surrogate data-sets series to build climate networks. This is obtained by shuffling full years in the time records of each site, keeping the order within each year un-changed. The correlations $R(H)$ of the network built by the surrogate data series is shown in figure 2.

[20] The second validation test is performed by comparing the correlation $R(H^*)$ to correlation with sequences that have the same dynamical behavior as the NAO-Index. For this purpose we use earlier data of NAO-Index from years 1823-2001 [URL2] to compute the extended NAO-indices:

$$I_y(Y), y = 1865,...,1945 \qquad (7)$$

The sub-index y is the first year of the surrogate sequence. This way we form 81 surrogate sequences (configurations) each with the same length (55 years) like the NAO-Index, $I(Y)$. Each sequence is shift by 1 year.

[21] We calculate the correlations between the number of the strong links, $\sigma_H(Y)$ computed in section, and each one of the surrogate sequences, $I_y^E(Y)$. We show in figure 4 the histogram of the correlation values, for each of the 4 data sets mention above. One can see that for all data sets the real correlation (mark with a red line) is significantly higher than all of the surrogate correlation values.

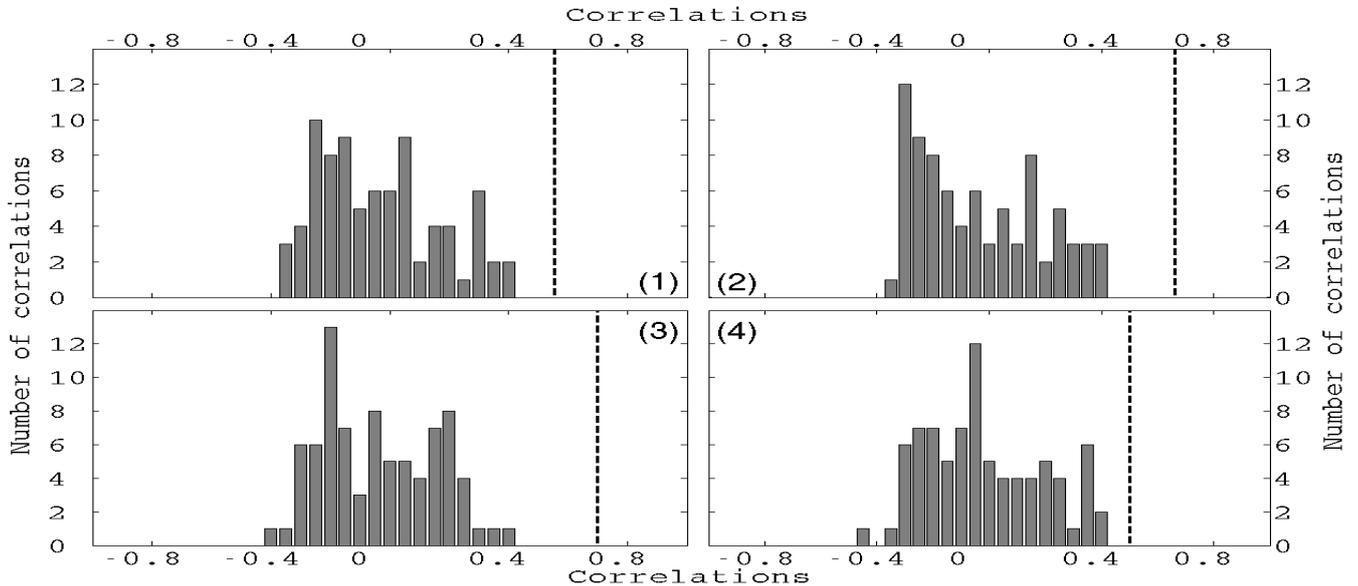

**Figure 4** The histograms (gray bar) of the correlation values between the number of the strong links, $\sigma_H(Y)$ (describe in [11]) and the surrogate indices, $I_y(Y)$ (described in [20]), and the correlation values $R(H^*)$ (black line, given in Table 1). The 4 panels describe the behavior of the network of 4 types of data sets: (1) Air temperature field at 500hPa. (2) Geopontential height field at 500hPa. (3) Air temperature field at 1000hPa. (4) Geopontential height field at 1000hPa.

## Summary and Conclusions

[22] We find that the NAO variations influence the number of strong links (high correlations) in the climate network. In particular, for negative index periods we observe the weakening of many strong links, which appear during positive index periods. We also find that there is a broad range of correlation values that is most sensitive to the NAO. Our results suggest that the links in the climate network are more sensitive to NAO variations compared to the temperature or pressure changes.